\documentclass[a4paper,aps,prd,showpacs,twocolumn, nofootinbib
]{revtex4-2}
\usepackage{blindtext}

\usepackage{graphicx}
\usepackage{dcolumn}
\usepackage{bm}
\usepackage[utf8]{inputenc}
\usepackage{amsmath}
\usepackage{amsfonts}
\usepackage{mathtools}
\usepackage{amssymb}
\usepackage{frontespizio}
\usepackage[T1]{fontenc}
\usepackage[utf8]{inputenc}
\usepackage{lmodern}
\usepackage[english]{babel}
\usepackage{latexsym}
\usepackage{braket}
\usepackage{physics}
\usepackage{slashed}
\usepackage{hyperref}
\usepackage{verbatim}
\usepackage{graphicx}
\usepackage{units}
\usepackage{tikz}
\usepackage[normalem]{ulem}
\usepackage{booktabs}
\usepackage{appendix}
\usepackage[textheight=\textheight,textwidth=\textwidth,bindingoffset=0cm,vcentering]{geometry}
\usepackage{color}
\def\laq{~\raise 0.4ex\hbox{$<$}\kern -0.8em\lower 0.62ex\hbox{$\sim$}~}
\def\gaq{~\raise 0.4ex\hbox{$>$}\kern -0.7em\lower 0.62ex\hbox{$\sim$}~}

\usepackage{titlesec}								

\DeclareMathOperator{\Erfc}{Erfc}

\def\beq{\begin{equation}}
	\def\eeq{\end{equation}}
\def\bea{\begin{eqnarray}}
	\def\eea{\end{eqnarray}}
\def \bes{
\begin{split}}
    \def \ees{\end{split}}

\def \non {\nonumber}

\def \le {\left}
\def \ri {\right}
\def \Rpert {\cal R}

\def \a {\alpha}

\def \b {\beta}

\def \g {\gamma}

\def \d {\delta}

\def \s {\sigma}

\def \r {\rho}

\def \om {\omega}
\def \Om {\Omega}
\def \f {\phi}

\def \t {\tau}

\def \mc {\mathcal}

\def \Mp {M_{\rm P}}

\def \b {\beta}
\def \a {\alpha}

\def \r {\rho}
\def \om {\omega}
\def \Om {\Omega}

\begin{document}
\title{Primordial black holes formation in a early matter dominated era from the pre-big bang scenario}

\author{Pietro Conzinu$^a$}
\email{pietro.conzinu@phd.unipi.it}
\author{Giovanni Marozzi$^a$}
\email{giovanni.marozzi@unipi.it}
\affiliation{$^{a}$Dipartimento di Fisica, Universit\`a di Pisa, Largo B. Pontecorvo 3, 56127 Pisa, 
Italy,\\
and Istituto Nazionale di Fisica Nucleare, Sezione di Pisa, Italy}

\begin{abstract}
We discuss the production of primordial black holes in an early matter dominated era, which typically takes place in string inspired early universe cosmological models. In particular, 
we consider a pre-big bang scenario (extending previous results regarding formation in the radiation dominated era) where the enhancement of curvature perturbations is induced by a variation of the sound-speed parameter $c_s$ during the string phase of high-curvature inflation. After imposing all relevant observational constraints, we find that the considered class of models is compatible with the production of a large amount of primordial
black holes, in the mass range relevant to dark matter, only for a small range of the parameters space. 
On the other hand, we find that a huge production of light primordial black holes may occur both in such matter dominated era and in the radiation dominated one.
\end{abstract}

\maketitle

\section{Introduction}
\label{Int}
Primordial black holes (PBHs) have attracted considerable attention in recent years and several different formation mechanisms were proposed (see e.g. \cite{Carr:2021bzv, Khlopov:2008qy, Escriva:2022duf}  for a review of the different proposals). 
PBHs may form due to the collapse of large fluctuations in the early universe. In particular, one can distinguish two particular cases in which the collapse happens in a radiation era and in a matter era. In the first case, the most studied in literature, due to the presence of radiation pressure, only large enough fluctuations can collapse. In the second case, instead, there is no pressure contribution and the effects of shape deformations become
crucial to define properly the collapse \cite{Harada:2016mhb, Harada:2017fjm}.
While in standard scenarios we have only one matter dominated era happening after the radiation dominated one, in many alternative scenarios there can be an early matter
dominated era before the radiation one, increasing the interest in studying the formation of PBHs in such era.
Here, we consider the possibility that PBHs may be form from the gravitational collapse of primordial density fluctuations in the early matter era that follows a pre-big bang scenario \cite{Gasperini:2002bn}, extending our previous work \cite{Conzinu:2020cke}, where we considered the formation for density fluctuations that re-enter the horizon in the radiation era.
The paper is organized as follows. In Sec.~\ref{Sec2} we describe the PBHs mass at formation for both matter and radiation domination era. In Sec.~\ref{Sec3} we introduce the PBHs abundance parameter, its relation with dark matter and with the primordial power spectrum in the above two eras. In Sec.~\ref{Sec4} we show the results obtained for formation in the early matter era that takes place for the pre-big bang scenario. Finally, we draw our conclusions in Sec.~\ref{Conc}.

\section{Primordial Black Holes Mass at Formation: Matter vs Radiation Era}
\label{Sec2}

If the density contrast $\d \equiv \d\r/\r$ is large enough (exceeding a critical value $\d_c$ \cite{Carr:2021bzv, Escriva:2022duf, Carr:1975qj}) then the fluctuations, when re-enter the horizon,  collapse directly in black holes.
The mass of a primordial black hole $M_{pbh}$, when it forms, is proportional to the mass contained in the Hubble horizon at the formation time $M_H=\frac{4\pi}{3}\r H^{-3}=4\pi M^2_p H^{-1}\,$ \cite{Carr:1974nx}.

Defining the re-entry time of a fluctuation with wave-number $k$ by $k=a(t_k)H(t_k)\equiv a_k H_k$, the PBH mass can be written as
\beq
M_{pbh}=4\pi \frac{M^2_p}{H_k}=4\pi \frac{M^2_p}{k}a_k \,.
\eeq 
In this way one can express directly the mass $M_{pbh}$ in terms of the wave number $k$ . \\
For matter dominated era, defining $t_d\sim H^{-1}_d$ as the time when the early matter era ends (and begin the radiation era), it holds~\footnote{Here $a_{eq}$ and $H_{eq}$ are the scale factor and the Hubble parameter at the radiation-matter equilibrium era.}
\beq
\begin{split}
\frac{a_k}{a_0}=&\le(\frac{a_k}{a_d}\ri)\le(\frac{a_d}{a_{eq}}\ri)\le(\frac{a_{eq}}{a_0}\ri)= \nonumber\\
=&\le(\frac{H_d}{H_k}\ri)^{2/3}\le(\frac{H_{eq}}{H_d}\ri)^{1/2}\le(\frac{H_0}{H_{eq}}\ri)^{2/3}\,,
\end{split}
\eeq
using the relation $H_k=k/a_k$ and setting $a_0=1$, we can solve in terms of $a_k$
\beq
a_k=\le(\frac{H_{eq}}{H_d}\ri)^{-{1\over2}}\le(\frac{H_0}{k}\ri)^2 \,,
\eeq
while in the case of radiation era, since
\beq
\begin{split}
&\frac{a_k}{a_0}=\le(\frac{a_k}{a_{eq}}\ri)\le(\frac{a_{eq}}{a_0}\ri)=\le(\frac{H_{eq}}{H_k}\ri)^{1/2}\le(\frac{H_0}{H_{eq}}\ri)^{2/3}\,,
\end{split}
\eeq
we obtain
\beq
a_k=\frac{H_{eq}}{k}\le(\frac{H_0}{H_{eq}}\ri)^{4/3} \,.
\eeq
Therefore, the PBHs mass can be written as
\bea
M_{pbh}\simeq &&4\pi\frac{M^2_pH_{eq}}{k^2}\le(\frac{H_0}{H_{eq}}\ri)^{4/3}, \quad ~~~\,(\textbf{RD})\,,\\
M_{pbh} \simeq &&4\pi \frac{M^2_p}{k^3}\le(\frac{H_{eq}}{H_d}\ri)^{-{1\over2}}H_0^2 \; ,\quad \quad (\textbf{MD})\,.
\eea
One can note that in the two abobe cases the dependence from $k$ is different and in the case of MD the mass depends also on the duration of the matter phase by the factor $H_d$. \\

\section{Abundance and Relation with Cold Dark Matter}
\label{Sec3}

In order to quantify the constraints on  PBHs, their abundance and their possible dark matter nature, one can consider two different parameters. The so-called energy density fraction $\beta$, that quantifies the abundance of PBHs at formation (see e.g. \cite{Carr:2021bzv,Carr:2020xqk,Carr:2020gox}),
\begin{equation}
	\beta\equiv \frac{\rho_{pbh}}{\rho_{tot}}\Bigg|_{ at formation}\,,
	\label{beta def}
\end{equation} 
where $\r_{pbh}$ and $ \r_{tot}$ are the density energy in form of PBHs and the total energy respectively. \\
And the parameter $f_{pbh}$, defined as (see e.g. \cite{Sasaki:2018dmp})
\beq
f_{pbh}\equiv \frac{\r_{pbh}}{\r_{cdm}}\Bigg|_{t_0}\,.
\label{fpb}
\eeq
which gives us the relative amount of cold dark matter in form of PBHs today, where $\rho_{cdm}$ is the dark matter contribution. Since  the majority of constraints for PBHs are usually given in terms of this parameter it is important to relate $f_{pbh}$ and $\b$ and then translate constraints from $f_{pbh}$ directly on $\b$ and, as we will explicitly see later, on constraints for the primordial power spectrum. \\
By definition, at the radiation-matter equilibrium era, after all the PBHs have been formed, we have that \cite{Carr:2020gox}
\beq
f_{pbh}=\frac{\b}{\Om_{cdm}}\Bigg|_{eq} \,.
\label{f eq}
\eeq
On the other hand, in the case of formation in radiation era, since $\r_{pbh}\sim a^{-3}$ and $\r_{rad}\sim a^{-4}$, then $\b\sim a$; while in matter era formation it holds $\b=\b_0\simeq const$. Thus we can manipulate equations \eqref{fpb} and \eqref{f eq} and obtain the relation between $f_{pbh}$ and $\b$ in these two cases.\\

{\bf Radiation era:}
as mentioned before, PBHs might form if the density contrast exceeds a critical value $\d_c$. 
Assuming a probability distribution for the density contrast  $P(\d)$ then $\b$ can be expressed as \cite{Carr:2020gox}
	\beq
	\b=\int_{\d_c}^{\infty}P(\d)d\d \,,
	\label{generic mass fraction}
	\eeq 
	where $\d_c$ is such critical density \cite{Carr:2020xqk}. If the probability distribution is Gaussian we have
	\beq
	P(\d)=\frac{1}{\sqrt{2\pi \s^2}} e^{-\frac{\d^2}{2\s^2}}\,,
	\eeq
	where $\s=\s(R)$ is the mass variance given by
	\beq
	\s=\int_0^\infty \tilde{W}(kR)\mathcal{P}_\d (k)\frac{dk}{k}\,,
	\eeq
where $\mc{P}_\d(k)$ is the primordial power spectrum of $\d$ at horizon entry and $\tilde{W}$ is a window function smoothing over the comoving scale $R \sim (a_k H_k)^{-1}$. \\
The PBHs mass fraction \eqref{generic mass fraction} assumes the simple form
	\beq
	\b=\frac{1}{\sqrt{2\pi \s^2}}\int^\infty_{\d_c}\exp{-\frac{\d^2}{2\s^2}} d\delta=\Erfc\left(\frac{\d_c}{\sqrt{2}\s}\right)\,,
	\label{beta rad}
	\eeq
	where $\Erfc$ is the complementary error function.

Combining Eq. \eqref{f eq} with entropy conservation, we get the connection between $\b$ and $f_{pbh}$
\beq
f_{pbh}=\b \frac{\Om_\g^0}{\Om^0_{cdm}}\frac{g(T_k)g_s(T_0)}{g(T_0)g_s(T_k)}\frac{T_k}{T_0} \,.
\eeq
where $\Om_\g^0$ is the radiation energy density today, and $g$, $g_s$ are the number of effective temperature and entropy relativistic degrees of freedom respectively. \\

{\bf Matter era:}
The situation is different if the formation happens in a matter dominated era. This case has been of less interest during the past years and only few studies have been done (see e.g. \cite{Harada:2004pe, Harada:2017fjm, Ballesteros:2019hus} ), essentially because in the standard scenario we have only one matter era happening after the radiation dominated era. As a consequence only PBHs of enormous mass could be formed during such era. 
On the other hand, in many scenarios, mostly coming from string theory, there can be a matter dominated era also before the radiation one, as it is the case of pre-big bang scenario \cite{Gasperini:2002bn}.  
 This gives new motivations to study the PBHs formation in a matter era and motivated this work. 
 
 While during the radiation era, 
 as a consequence 
 of the presence of radiation pressure, one obtain a critical density $\d_c$ of order unity, beyond which one obtains a collapse of the overdense region, in a matter contest this not holds and the effects of shape deformation are crucial to define properly the critical density.
Following \cite{Harada:2016mhb}, the criterion for formation in matter era is given in terms of the hoop conjecture~\footnote{The hoop conjecture \cite{Misner:1974qy} states that a black hole forms when a mass $M$ collapse into a region whose circumferences in every direction are smaller than $\frac{4\pi G M}{c^2}$.\\ \\
}. By numerical analysis it was shown in \cite{Harada:2016mhb} that holds
	\beq
	\b_0\sim 0.056\s^5\,.
	\label{beta matter}
	\eeq
  Furthermore, in \cite{Harada:2017fjm} it was pointed out that if one takes into account the angular momentum the expression given before holds only for $\s>\s_{ang}=0.005$, while for smaller values was derived the semi-analytic expression 
		\beq 
		\b_0\sim 1.321 \times 10^{-7} f_q( q_c) \mc{I}^2 \s^2 \exp{\le(-0.1474 \frac{\mc{I}^{4/3}}{\s^{2/3}}\ri)} \,,
		\label{beta matter angular}
		\eeq
where $\mc{I}$ is related to the variance of the angular momentum and $f_q(q_c)$ is the fraction of mass with quadrupole asphericity $q\ll q_c \sim 2.4(\mc{I}\s)^{1/3}$, both $\mc{I}$ and $f_q$ are parameters of order unity.

In order to connect the two parameters $\b$ and $f_{pbh}$, we assume an instantaneously matter-radiation transition, such that
\beq
3M^2_{P}H_d^2=\r_d=\frac{\pi^2 g(T_d)}{30}T_d^4\,,
\eeq
where as before the subscript $"d"$ states for the end of matter era and $T_d$ is the correspondent temperature, finally $g(T_d)$ counts the effective number of degrees of freedom.   Then, using again entropy conservation, we obtain
\beq
f_{pbh}=\b_0 \frac{\Om_\g^0}{\Om^0_{cdm}}\frac{g(T_d)g_s(T_0)}{g(T_0)g_s(T_d)}\frac{T_d}{T_0} \,.
\eeq

For a rough estimation we can assume $g\simeq g_s$ for all the epochs and obtain
\bea
f^{RD}_{pbh}=\b \frac{\Om_\g^0}{\Om^0_{cdm}}\frac{T_k}{T_0}\,,
\label{fpbhRD}\\
f^{MD}_{pbh}=\b_0 \frac{\Om_\g^0}{\Om^0_{cdm}}\frac{T_d}{T_0}\,. \label{f in matter}
\eea
where we have $\Om^0_{cdm}\simeq 0.26, \, \Om_\g^0\simeq 10^{-4}, T_0\simeq 2.7 ~\unit{K}$.
In Eq.(\ref{fpbhRD}) we can make explicit the PBH mass dependence by the relation $T_k=( 4\pi M_p/M_{pbh})^{1/2}(3/g(T_k))^{1/4}$.\\

As pointed out above, if formation happens in the radiation era $\beta \sim a$, while in matter era we have that $\b=\beta_0$ is nearly constant. Thus, we can relate Eqs. \eqref{beta rad} and \eqref{beta matter} by the following simple relation
\beq
\b_0 \simeq \le(\frac{H_k}{H_d}\ri)^{2/3}\b\,.
\eeq
Using this result, all relations between the parameter $\b$ and $f$ obtained for the radiation dominated era can be then directly converted to the case of formation in an early matter era.\\

\subsection{Relation with Primordial Power Spectrum}
\label{subsec 5.5.1}
Starting from the above results, we can now translate the constraints on PBHs abundance in terms of constraints on the correspondent amplitude of the comoving curvature perturbation $\mathcal{R}$. \\
The density contrast $\d$ is related to $\mathcal{R}$ by \cite{Ballesteros:2019hus}
\beq
\d=\frac{2(1+\om)}{5+3\om}\mathcal{R}\,,
\eeq
 where $\omega$ is the equation of state parameter ($\omega=0$ for matter era and $\omega=1/3$ for radiation era).
Then, since the variance $\s^2\sim P_\d$, we have
\begin{subequations}
\begin{align}
\s^2\sim \frac{16}{81}\mc{P_R} \qquad \text{RD} \,, \nonumber \\
\s^2\sim \frac{4}{25}\mc{P_R}  \qquad \text{MD} \,. \nonumber
\end{align}
\end{subequations}

We can now invert Eqs. \eqref{beta rad} and \eqref{beta matter} to obtain, for a radiation dominated era
\beq\label{eq33}
P_{\d}\sim \s^2= \le(\frac{\d_c}{\sqrt{2}\Erfc^{-1}(\b)}\ri)^2 \,,\qquad 
\eeq
while for a matter dominated era we have
\begin{subequations}\label{eq34a}
\begin{align}
P_{\d}\sim \s^2 = \le(\frac{\b_0}{0.056}\ri)^{2/5}\,, \qquad  \s > 0.005\,,\\
P_{\d}\sim \s^2=10^{-4}W \left(\frac{0.05}{\b^{1/3}_0}\right)^{-3}\,, \qquad  \s < 0.005\,,
\end{align}
\end{subequations}
where $W$ is the Lambert $W$-function.\\

Requiring to have the total or at least an important fraction of cold dark matter in form of PBHs, one obtains important constraints on the needed value of the primordial power spectrum. 
Comparing Eqs. \eqref{eq33} and \eqref{eq34a} it is then clear the main difference between formation in radiation dominated and matter dominated eras: in the former the PBHs abundance is exponentially sensitive to the primordial power spectrum $\mathcal{P_R}$, while in the second case the dependence is only a power-law. 

Furthermore, in the case of matter era, it should be noted that in order to have collapse we require that $\d\sim 1$ and thus only the perturbations which have time to grow until such value can collapse. Since the linear density contrast grows as the scale factor during a matter dominated era, only fluctuations that reach non-linearity before the end of the early matter era could collapse into a black holes, i.e. only fluctuations for which $\s>\s_{nl}=(H_d/H_k)^{2/3} $ could really collapse to form a PBH \cite{Ballesteros:2019hus}.

\section{Pre-Big Bang}

We consider the production of PBHs in the pre-big bang scenario, assuming the class of models discussed in  \cite{Conzinu:2020cke}. 
Such a class of models follows a low-energy effective string action for the massless modes (graviton, dilaton $\f$ and the Kalb-Ramond axion field $\chi$). In this scenario the accelerated evolution, driven by the dilaton field, starts from a low-energy phase corresponding to the string perturbative vacuum at $\tau= - \infty$. 
For this case, one has an initial Kalb-Ramond axion background trivial, $\chi = 0$, but with quantum fluctuations
 $\delta \chi$ non-vanishing.
This first dilaton-driven inflationary phase ends at the epoch $ \tau = - \tau_s$, when the S-frame background
curvature reaches the string scale. At that point, to have a consistent description of the dynamics we need to include high-
curvature $\a'$ corrections into the string effective action, i.e. we are entering in the \textit{string phase}.
During this second high-energy inflationary phase, from $ \tau = -\tau_s$ to $ \tau =-\tau_1$, the S-frame curvature $H$ tends
to stay constant, while the effective string coupling keeps growing from $g_s \equiv g(-\tau_s)$ to $g_1 \equiv g(-\tau_1)$ \cite{Gasperini:1996fu}. Such a growth can be parameterized by a simple power-law behavior in conformal time
as $g(\tau) \sim (-\tau)^{-\a}$, where $\a$ is an unknown (positive) parameter.
This second regime of inflationary pre-big bang evolution should ends at the time $\tau =-\tau_1$,
where one would expect a graceful-exit from pre- to post-big bang universe \cite{Gasperini:2002bn,Gasperini:2007zz}.  

The scalar perturbations that exit during the string phase are affected by the higher-order $\alpha \prime$ corrections. 
The contribution of such corrections may produce, in general, an effective value $c_s \not=1$ (and a consequent modification of the spectra) for all types of background fluctuations \cite{Gasperini:2002bn,Gasperini:2007zz,Gasperini:1997up}. 
In our context, in particular, this happens for the scalar perturbations (sourced by the dilaton), and for the perturbations of the Kalb-Ramond axion field. The presence of this last type of fluctuations is crucial for the production, via the curvaton mechanism \cite{Lyth:2001nq}, of a viable spectrum of adiabatic curvature perturbations \cite{Bozza:2002fp, Bozza:2002ad}.
Thus, the study of perturbations in this scenario considering also the string phase, can be schematically described, at the linear order, by a Mukhanov-Sasaki-like equation 
\beq
v'' - \left( c_s^2 \nabla^2 + {z''\over z} \right) v=0\,
\label{Mukhanov eq.}
\eeq
where the prime means derivation wrt conformal time $\tau$, $v$ is the canonical variable, that diagonalizes the perturbed action, defined  as $v=z R$, with $R$ the curvature perturbation and $z $ the so-called pump field, and $c_s$ plays the role of an effective sound speed that  emerges when one considers the perturbations in the string phase. 

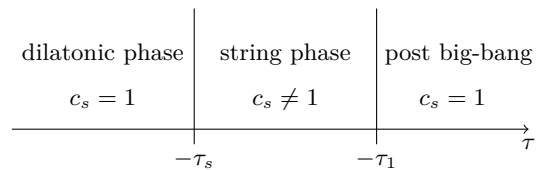
\begin{figure}[h!]
\centering
\begin{tikzpicture}[scale=0.4]
   \draw[->] (-9,0)-- (8,0);

 \draw (-6,2.5) node {\small{dilatonic phase}};

    \draw (-6,1) node {$c_s=1$};

    \draw (-3,-0.5)-- (-3,4);

    \draw (-3,-1) node {$-\tau_s$};

\draw (0,2.5) node {\small{string phase}};

    \draw (0,1) node {$c_s \neq 1$};

    \draw (+3,-0.5)-- (+3,4);

    \draw (+3, -1) node {$-\tau_1$};
    
\draw (5.7,2.5) node {\small{post big-bang}};

    \draw (5.5,1) node {$c_s=1$};
     \draw (8, -0.5) node {$\tau$};

\end{tikzpicture}
\vspace{-.2 cm}
\caption{Sketch of three phases} \label{Sketch}
\end{figure}

We have considered the perturbations for three phases, a initial dilaton-driven low energy phase, a second string phase and a third standard post-big bang phase, as reported in figure \ref{Sketch}.\\

We then solve Eq. \eqref{Mukhanov eq.} in these three phases separately, obtaining a normalized solution for the scalar perturbation ${{{\Rpert}}}_{k}=v_{k}/z$ \cite{Conzinu:2020cke}.
For our purpose, we are interested in the primordial amplitude of the perturbation modes when re-entering the horizon after the two phases of accelerated evolution. More precisely, we want to evaluate the dimensionless power spectra
\beq
\mc{P_{\Rpert}}(k) = {k^3\over 2 \pi^2} \left| {{\Rpert}}_k^3\right|^2_{|k\tau|=1},
\label{29}
\eeq
for $\tau > -\tau_1$.
We then obtain the following spectra for the curvature perturbation sourced by the dilaton $\f$ and axion $\chi$ \cite{Conzinu:2020cke}:
\\

\begin{widetext}
\begin{equation}
\hspace{-0.8cm}
\begin{minipage}{0.5\textwidth}
{\hspace{0.8cm}\bf{Axion spectrum}}
\begin{align*}
\mc{P}_\mc{R}^\chi(\om) ~&\simeq ~ {f^2\over 2\pi^2}\left(H_1\over \Mp\right)^2 
\left(\om\over \om_1\right)^{3-|3+2\a|} c_\chi^{-1-|3+2\a|}~,\non
 \\ &\simeq~
{f^2\over 2\pi^2 } \left(H_1\over \Mp\right)^2 
\left(\om_s\over \om_1\right)^{3-|3+2\a|}\left(\om\over \om_s\right)^4\,~~~~,\non
\\ &\simeq~
{f^2\over 2\pi^2 } \left(H_1\over \Mp\right)^2 
\left(\om_s\over \om_1\right)^{3-|3+2\a|}\left(\om\over \om_s\right)^{n_s-1},\non
\end{align*} \quad
\end{minipage}
\hspace{-0.3cm}
\begin{minipage}{0.5\textwidth}
{\hspace{0.8cm}\bf{Dilaton spectrum}}
\begin{align}
\mc{P}_\mc{R}^\f(\om) ~&\simeq~  {1\over 2\pi^2 }\left(H_1\over \Mp\right)^2 
\left(\om\over \om_1\right)^{3-|3-2\a|} c_\f^{-1-|3-2\a|}&&,
&& {\om_s\over c_\f}<\om<{\om_1\over c_\f},\non \\
 &~\simeq~
{1\over 2\pi^2} \left(H_1\over \Mp\right)^2 
\left(\om_s\over \om_1\right)^{3-|3-2\a|}\left(\om\over \om_s\right)^4&&,
&& \om_s<\om<{\om_s\over c_\f}, \non
\\ &~\simeq~
{1\over 2\pi^2} \left(H_1\over \Mp\right)^2 
\left(\om_s\over \om_1\right)^{3-|3-2\a|}\left(\om\over \om_s\right)^{3}&&,
&&~~~~\om<\om_s. \non
\end{align}
\end{minipage}
\end{equation}
\end{widetext}

Where $f^2 \simeq 0.137$ is the transfer function that relates the axion modes to the curvature perturbations, $\om_s$ and $\om_1$ are the frequency scales at the beginning and at the end of the string phase, $\a$ parameterize the evolution in the string phase~\footnote{Note that we have changed for convenience the notation for the axion field $\chi$ and the parameter $\a$ with respect to \cite{Conzinu:2020cke}, the two notations are mapped by $\chi \rightarrow \s$ and $\a \rightarrow \b$.}, and $H_1$ is the Hubble parameter at the end of the string phase \cite{Conzinu:2020cke, Gasperini:2007zz}.
We can then consider a 2-dimensional parameter space in terms of the redshift parameter $z_s$ and of the
the overall growth of the effective coupling $g$ during the string phase as:
\beq
 z_s= {\t_s\over\t_1} = {\om_1 \over \om_s},\,\qquad \qquad {g_s \over g_1} = \le({\t_s \over \t_1}\ri)^{-\a} = z_s^{-\a}. \nonumber
\label{parameters}
\eeq
Finally, for the model under consideration we should impose the appropriate
constraints arising from phenomenological as well as self-consistency conditions, in order to
explore the allowed region of parameter space \cite{Gasperini:2017fqw, Gasperini:2016gre, Conzinu:2020cke}.
In particular we impose the stability of the background, a growing string coupling, and a sub-Planckian curvature. We also impose the CMB constraints at the pivot scale $k_*= 0.05 \,,\unit{Mpc}^{-1}$. Finally, to do not interfere with the
standard nucleosynthesis scenario via a dust-dominated phase, we have to impose that the
decay of the oscillating axion occurs at a scale preceding the nucleosynthesis. Furthermore, we also assume that the axion-dominated
phase is short enough to affect only the highest frequency branch of the spectra. 
We refer the interested reader to \cite{Conzinu:2020cke} for more details.



\section{Formation in the early matter era}
\label{Sec4}

For the early matter era, using Eq.\eqref{f in matter} and inserting the numerical values, we obtain the following dark matter abundance
\beq \label{eq.27}
f^{MD}_{pbh}\sim \le(\frac{\b_0}{5.5\times10^{-15}}\ri)\le(\frac{T_d}{10^5 GeV}\ri)\,.
\eeq

Following \cite{Conzinu:2020cke} we have ${H_d\over M_P}=({H_1\over M_P})^3$  and $T_d \sim (H_d M_p)^{1/2}$,
thus requiring $f_{pbh}\sim 1$ in Eq. \eqref{eq.27} we obtain 
\beq
\b_0 \sim 5 \times 10^{-29} \le({H_1\over M_p}\ri)^{-3/2}\,,
\eeq
where we have used $\s^2 = {4\over 25}P_\mc{R}$.\\
Being interested to the implications that PBHs formation has for dark matter, one can consider firstly PBHs with mass $M_{pbh} \sim 10^{18}-10^{22} \unit{g}$ \cite{Escriva:2022duf, Carr:2020gox}. 
On the other hand, if one not assume instantaneous collapse, since the density contrast grows as the scale factor during matter era, one has that (as mentioned) only fluctuations with $\s > \s_{nl}= (H_d/H_k)^{2/3}$ reach non-linearity during the matter dominated era, while fluctuations with $\s < \s_{nl}$ will not collapse. 

\subsection{Collapse in matter dominated era neglecting $\s_{nl}$}
As described above, in the case of formation in matter era we have two different relations for the abundance of PBHs. Firstly, neglecting the evolution of fluctuations during the matter era (i.e. neglecting  $\s_{nl}$), we can consider the two cases:

\begin{itemize}
\item Case A: $\s > \s_{ang}$ \label{case a}.

Requiring all dark matter in form of PBHs, we obtain
\beq
P_{\mc{R}} > 10^{-10} \le({H_1\over M_p}\ri)^{-3/5}\, ,
\eeq

\item Case B: $\s < \s_{ang}$.

Requiring all dark matter in form of PBHs, we obtain
\beq
P_\mc{R} > \frac{7\times 10^{-4}}{W\left(6\times 10^5 \left(\frac{H_1}{M_p}\right)^{1/2}\right)^3} \, ,
\eeq

\end{itemize}

\subsection{Collapse in matter dominated era taking into account non-linearity}
Now we take into account the evolution of fluctuations during the matter era.
When $\s > \s_{ang}$, the rotation of the collapsing region can be neglected, and the reach of linearity is expressed by $\s > \s_{nl}$; while when $\s < \s_{ang}$ the angular momentum becomes important and the constraint is stronger \cite{Ballesteros:2019hus}.
As a consequence, we obtain the following constraints, that should be added to those obtained from $\b$ 
\vspace{-1mm}
\begin{subequations}
\begin{align}
\s > \s_{nl}&= \le({H_d\over H_k}\ri)^{2/3} \, , ~\qquad \s > \s_{ang}\,, \\
\s > \s_{nl}&= \le(5{H_d\over 2 \mc{I} H_k}\ri) \, , \qquad \s < \s_{ang}\,.
\end{align}
\end{subequations}
We finally obtain the following cases:
\begin{itemize}
\item  Case A:  $\s > \s_{ang}$.

With respect to the previous case A we obtain the new condition
\beq
 P_{\mc{R}} > {25\over 4} \le({M_{pbh}\over M_p}\ri)^{4/3} \le({H_1\over M_p}\ri)^{4}\,.
\eeq

\item Case B: $\s < \s_{ang}$.

In this case, we have the new condition for the power spectrum

\beq
 P_{\mc{R}} > \le({25\over 4}\ri)^2 \le({M_{pbh}\over M_p}\ri)^{2} \le({H_1\over M_p}\ri)^{6}\,.
\eeq
\end{itemize}

\begin{widetext}
	
\begin{figure*}[h!]
\begin{minipage}{0.45\textwidth}
\hspace{-3 cm}
 \textbf{Production in matter era neglecting\\ non-linear evolution}\par\medskip
\hspace{-3 cm}
	\begin{minipage}{0.45\textwidth}
		\includegraphics[scale=0.35]{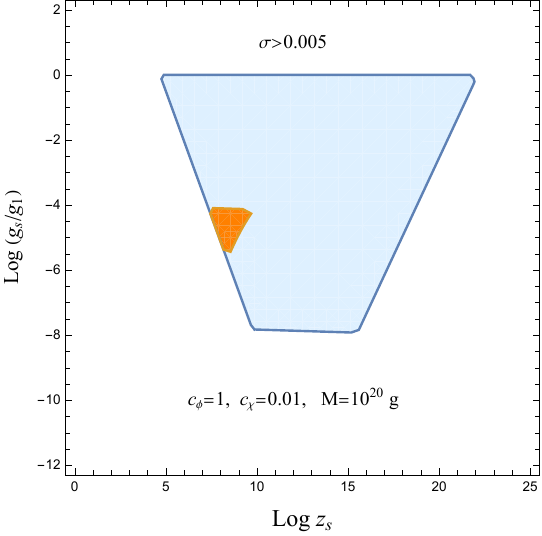}
	\end{minipage}
	\begin{minipage}{0.45\textwidth}
		\includegraphics[scale=0.35]{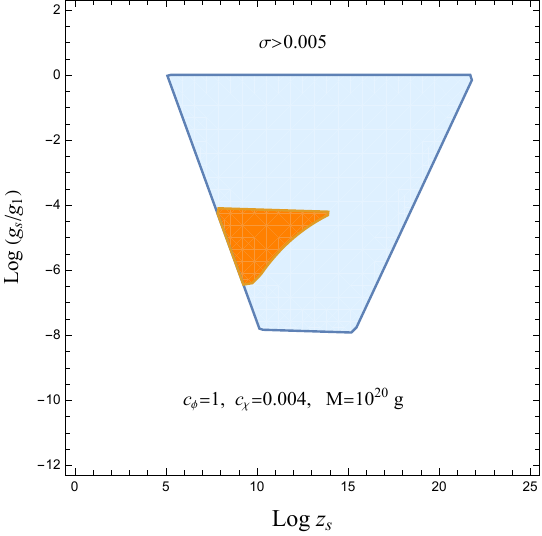}
	\end{minipage}

\hspace{-3 cm}
	\begin{minipage}{0.45\textwidth}
		\includegraphics[scale=0.35]{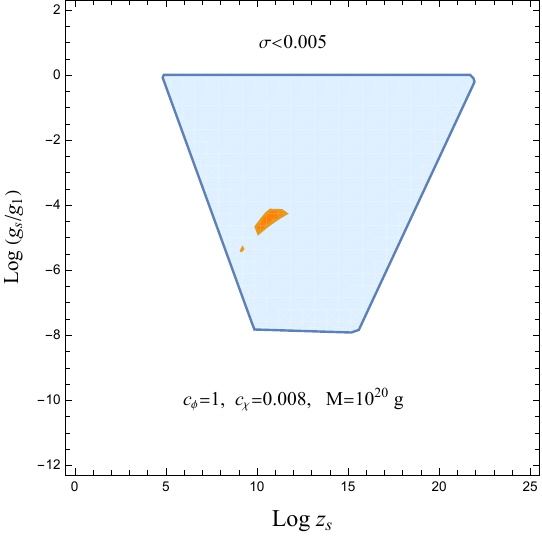}
	\end{minipage}
	\begin{minipage}{0.45\textwidth}
		
		\includegraphics[scale=0.35]{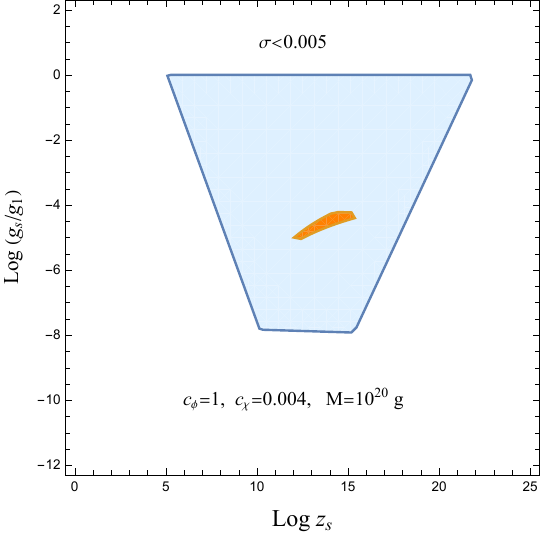}
	\end{minipage}

	\end{minipage}
	\begin{minipage}{0.45\textwidth}
\textbf{Production in matter era with non-linear evolution}\par\medskip
	\begin{minipage}{0.45\textwidth}
		\includegraphics[scale=0.35]{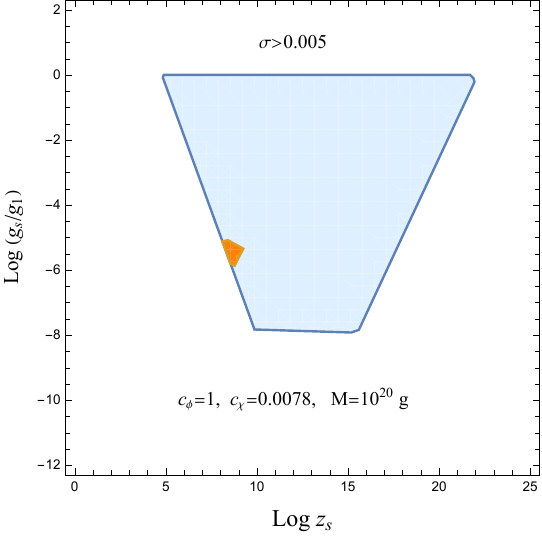}
	\end{minipage}
	\begin{minipage}{0.45\textwidth}
		
		\includegraphics[scale=0.35]{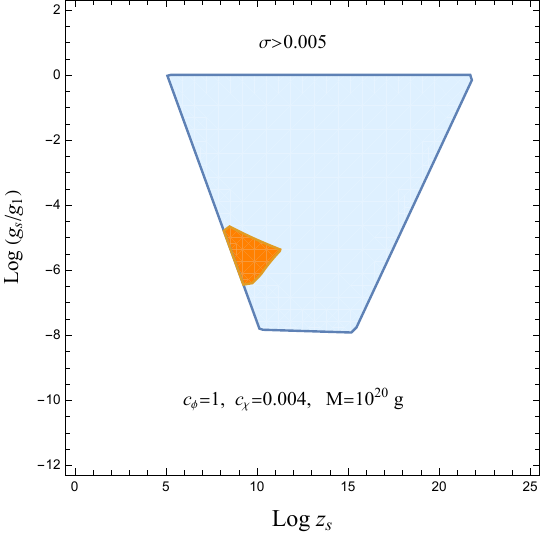}
	\end{minipage}
	\begin{minipage}{0.45\textwidth}
		\includegraphics[scale=0.35]{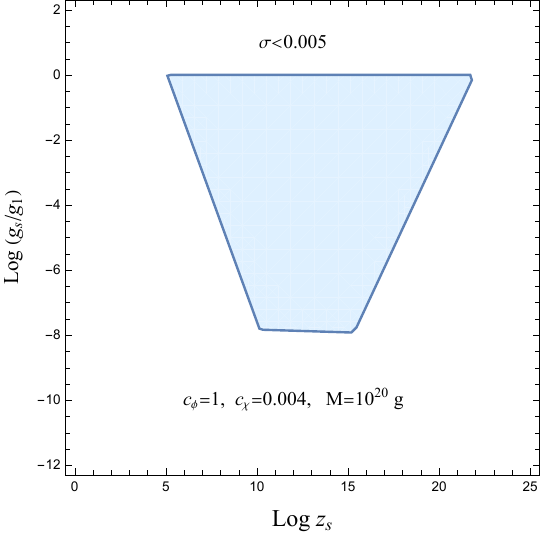}
	\end{minipage}
	\begin{minipage}{0.45\textwidth}
		
		\includegraphics[scale=0.35]{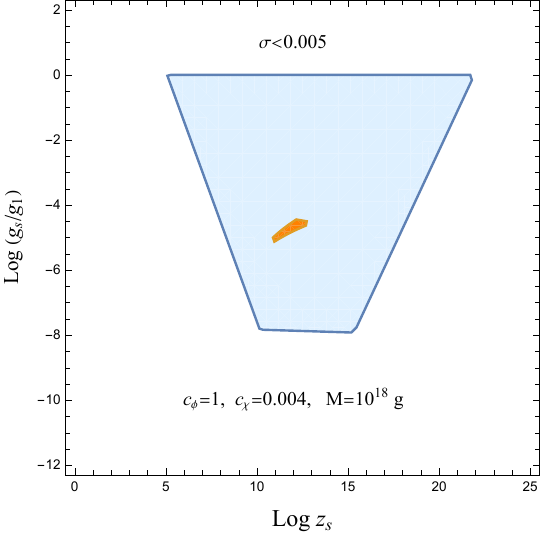}
	\end{minipage}

\end{minipage}
	\caption{   We show the production of PBHs (in orange) in matter era with and without the non-linear evolution (in the right and in the left panel respectively) for the case A (top) and B (bottom) for a mass of $10^{20} \unit{g}$ at varying axion sound speed (in orange). The parameter space (in light blue) is given in terms of $z_s= \tau_s/\tau_1$ and $g_s/g_1=z_s^{-\a}$. }
	\label{fig matter}
\end{figure*}

\end{widetext}

The results above are shown in Fig.~\ref{fig matter} where we show the parameter space (in blue) obtained by the typical constraints of the particular model of pre-big bang chosen \cite{Conzinu:2020cke, Gasperini:2016gre} and the region of parameter space (in orange) compatible with a production of PBHs that gives all the dark matter ($f \sim 1$). We note that there is not a huge superposition, in particular for the case b ($\s<0.005$.)

\subsection{Issues with too-light-PBHs}
\label{Sec5}
\vspace{-1mm}
One of the main constraints for PBHs comes from Hawking radiation \cite{Carr:2020gox}, in particular PBHs with $m<10^{15}\unit{g}$ should already be evaporated, with correspondent observable signatures\footnote{See \cite{Papanikolaou:2020qtd} for their conection with the production of gravitational waves.}. 
As a consequence, those light PBHs cannot be associated to dark matter. Moreover, strong constraints are given by
the effects that the evaporated particles could have on the big-bang nucleosynthesis \cite{Escriva:2022duf, Carr:2020gox}. So, one should check if in the context of the pre-big bang scenario the production of light PBHs is too efficient. Actually this seem the case, at least in a first analysis. Indeed, let us follow the same analysis above for light masses $10^{10}<m<10^{19} \unit{g}$. We show both the case of formation in radiation era Fig.\ref{Fig light_production in rad} and in matter era Fig.\ref{Fig light_production in mat}. We can see that, in particular for the case of matter era, there is a huge superposition in the parameters space that grows for smaller masses. However, such issue can be addressed in many ways. One for all, in a more realistic scenario, the sound speed depends on the modes and it changes differently for each mode considered. We then have that the formation of lighter PBHs is related to high frequency modes that exit from the horizon closer to the end of the string phase. During the end of such string phase other corrections (even non-perturbative ones) should be taken into account to produce a smooth transition. Such corrections can change drastically the sound speed, possible toward the standard unitary value, stopping the productions of PBHs and therefore addressing the above issue.
A detailed study of the properties of the production of such light PBHs would be really interesting.
In particular for understanding the related constraints and impact of such production, as the ones associated to PBHs which evaporated before big-bang nucleosynthesis. These can in fact affect the generation of baryon asymmetry (see e.g. \cite{Baumann:2007yr}). We postpone this study to a future work.

\begin{widetext}

\begin{figure*}[h!]
\begin{center}
\hspace{-2cm}
\textbf{Production of light PBHs in radiation epoch}\par\medskip
\end{center}
\includegraphics[scale=1]{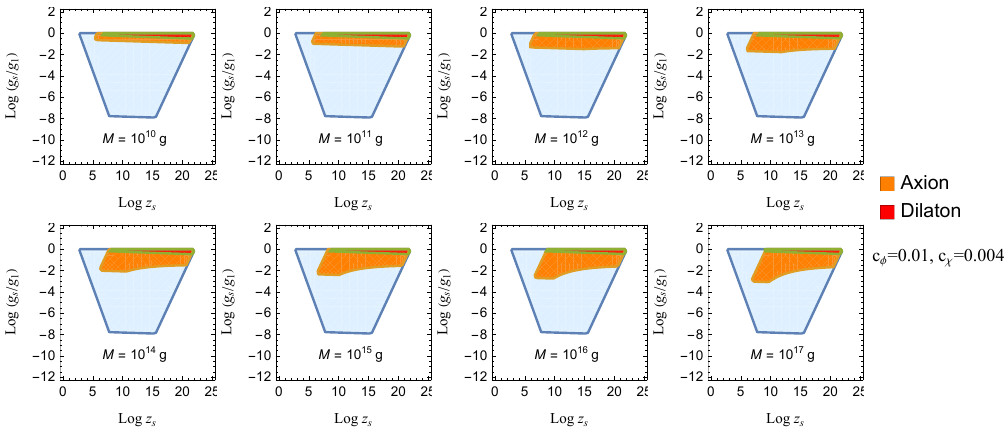}

\caption{We show the production of light PBHs at varying mass (mass expressed in grams) in the radiation era.  The parameters space (in light blue) is given in terms of $z_s= \tau_s/\tau_1$ and $g_s/g_1=z_s^{-\a}$.}
\label{Fig light_production in rad}
\end{figure*}

\begin{figure*}[h!]
\begin{center}
\textbf{Production of light PBHs in matter epoch}\par\medskip
\end{center}
\includegraphics[scale=1]{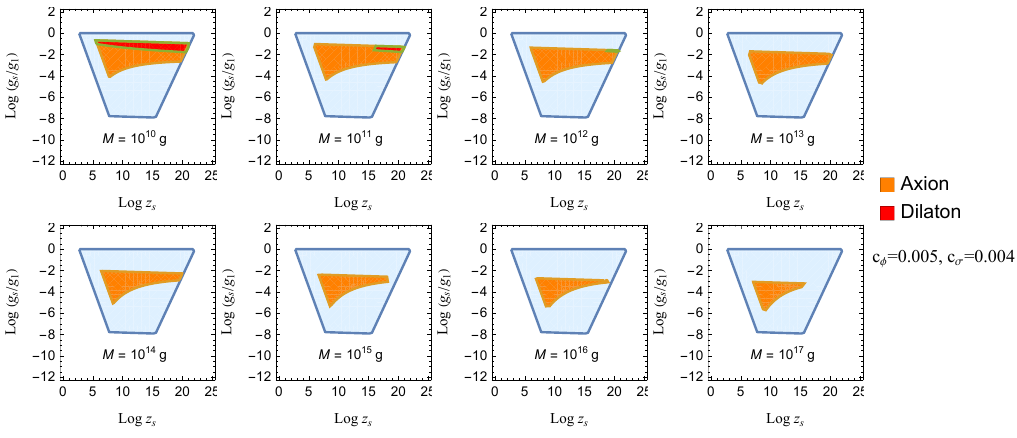}
\caption{We show the production of light PBHs at varying mass (mass expressed in grams) in the matter era. The parameter space (in light blue) is given in terms of $z_s= \tau_s/\tau_1$ 
and $g_s/g_1=z_s^{-\a}$.}

\label{Fig light_production in mat}
\end{figure*}
\vspace{20mm}

\end{widetext}

\section{Conclusions}
\label{Conc}
\vspace{-1mm}
In this paper we show how the production of PBHs, in the pre-big bang scenario, described previously in \cite{Conzinu:2020cke}, can be extended also to the case of formation in a early matter dominated era. This early matter era it is needed in order to produce the right amount of scalar perturbations by the curvaton mechanism. We show that in this case there is not a huge production of PBHs in the proper range of mass needed to explain dark matter (see Fig.\ref{fig matter}). Such a region is quite
sensitive to the effective value of the sound-speed parameter when $c_s \ll 1$. As showed in \cite{Conzinu:2020cke}, the superposition disappears for $c_s<0.033$ due to the shrinking of the parameter space allowed (in blue in the figure).
However, it seems that a large production of light PBHs ($m< 10^{18} g$) can happens both in the matter and radiation dominated eras (see Figs. \ref{Fig light_production in rad} and \ref{Fig light_production in mat}). As already discussed, this last eventuality could be avoided in several ways, for example adding new higher-order corrections (even non-perturbative) or with a more realistic model of the sound speed variation, in which such variation depends on the particular mode considered. We postpone this analysis to future works.

\section*{Acknowledgements}
We are very thankful to Maurizio Gasperini for useful discussions and comments, and for feedback on the manuscript.
We are supported in part by INFN under the program TAsP ({\it Theoretical Astroparticle Physics}).

\end{document}